# Determination of signal-to-noise ratio on the base of information-entropic analysis

Zhanabaev Z. Zh., Akhtanov S.N., Kozhagulov E.T., Karibayev B.A.

In this paper we suggest a new algorithm for determination of signal-to-noise ratio (SNR). SNR is a quantitative measure widely used in science and engineering. Generally, methods for determination of SNR are based on using of experimentally defined power of noise level, or some conditional noise criterion which can be specified for signal processing. In the present work we describe method for determination of SNR of chaotic and stochastic signals at unknown power levels of signal and noise. For this aim we use information as difference between unconditional and conditional entropy. Our theoretical results are confirmed by results of analysis of signals which can be described by nonlinear maps and presented as overlapping of harmonic and stochastic signals.

**Introduction**

The main characteristics of communication electronic systems are SNR and bit error rate (BER) defined via SNR [1].

As usual, SNR can be defined as signal power to noise power ratio [2-6]. This approach is developed in many works. For example, [7] contains a description of using of multiple linear regression with coefficients chosen for different types of noise for defining of SNR.

Estimation of SNR can be made via relation between autocorrelation functions of a signal and noise shifted by time [8], but noise and signal levels have arbitrary chosen values. Authors of [9] use wavelet signal filtering with certain chosen coefficients, and define SNR as relation between variance of signal and variance of noise. It was specified that in this case the SNR calculation is relatively fast. However, due to the limited bandwidth of wavelet functions spectrum, this method is not applicable to analyze very noisy signals. Effective SNR estimation defined as difference between standard SNR and signal quality indicator which also must be given previously is described in [10]. The single distinction between modified segmental SNR and standard SNR is in the fact that for calculation of modified SNR it is necessary to summarize SNR for separated time intervals [11].

SNR is used in many areas of science and engineering such as in wireless telecommunications systems [12-15], medicine [16-17], nuclear physics [18], neuroscience [19], sound technique [20-22], optoelectronics [23-24], nanotechnology [25-26], astrophysics [27-28], etc.

However, generally accepted methods and original algorithms for calculation of SNR used in the mentioned above papers have the following limitations:

- necessity to set value of noise level according to experimental details or conditional criteria;

- absence of standard algorithms for SNR definition;

- absence of a universal theoretical approach for SNR definition for signals with unknown noise level.

So, the problem follows from the described above: is it possible to define SNR as a relation between information and entropy? We formulate this problem by the following way because information is a universal measure of determinacy of a signal, and entropy is a measure of its uncertainty (noise). For solving of this problem we can accept that information is not a local characteristic, but an averaged value defining via difference between unconditional and conditional entropy [29]. Information entropy is often used in different research. For example, comparison of composite, refined and multi-scale cross-sample entropy of complex signals is described in [30]. Entropic analysis can be applied for the description of such complex signals as multi-fractal signals, financial time series, etc. Informational entropy can be used for signal filtering [31]. Entropy can be also applied for classification of infrasound signals [32]. Cognitive state of human subjects on the base of entropic analysis of physiological signal is described in [33].

Relationship between SNR and entropy is specified in [34], in this work value of maximum entropy of probability density function for convolutional noise is used for the description of modulation and SNR. An image filtering based on calculation of entropy is suggested in [35]. Seismic signal filtering based on wavelet transform and application of Shannon and Tsallis entropy for determination of SNR is described in [36]. Entropy analysis can be useful for the description of dynamical systems with chaotic behavior [37-38]. Unfortunately, in spite of the fact that relation between entropy and SNR has been described in these works, a ratio between information and entropy hasn't been used, and described above limitations for definition of SNR remain valid.

The aim of this work is to define value of SNR as information to entropy ratio (IER) for various signals (mixture of harmonic signal and noise, chaotic signals from dynamical systems [39-41]).

1. **A new algorithm for determination of SNR**

By the definition given in [29] value of full information of a signal $x = x(t)$ transmitted over a communication channel can be define as difference between one-dimensional Shannon entropy and conditional entropy as

$$I(x,y) = S(x) - S(x\,|\,y), \qquad (1)$$

where $y(t)$ is characteristic of a receiver. Unconditional Shannon entropy can be defined as

$$S(x) = -\sum_{i=1}^{N} p_i \ln(p_i), \qquad (2)$$

where $p_i$ is a probability of detecting of variable x in a i-th cell characterized by size δ, $S(x\,|\,y)$ is conditional entropy given as

$$S(x\,|\,y) = -\sum_{i=1}^{N}\sum_{j=1}^{M} P(x_i, y_j) \ln(P(x_i\,|\,y_j)). \qquad (3)$$

Here $P(x_i\,|\,y_j)$ is conditional probability. Defining of information according to Eq. (1) is possible if we have an empirical set of probabilities for time series of x(t) and y(t). For the description of dynamical systems we can accept $y(t) = x'(t)$, so, we consider the derivative of x(t) as a second variable. Instead of one-dimensional Shannon entropy $S(x)$ we use a two-dimensional full entropy $S(x,y)$. So, we can rewrite Eq.(1) as

$$\begin{cases} I(x,y) = S(x,y) - S(x|y) > 0, \\ S(x,y) = -\sum_{i=1}^{N}\sum_{j=1}^{M} P_{ij} \ln(P_{ij}), \end{cases} \quad (4)$$

where $P_{ij}$ is probability of detecting of a point in a cell of phase space (x,y).

Values of two-dimensional information and conditional entropy can be normalized to full entropy according to the following relations:

$$\tilde{I} = I(x,y)/S(x,y), \quad (5)$$

$$\tilde{S} = S(x|y)/S(x,y). \quad (6)$$

We determine value of IER as ratio of normalized information and conditional entropy as

$$IER = \tilde{I}/\tilde{S} = 1/\tilde{S} - 1 \quad (7)$$

Eq. (7) defines SNR with unknown noise level if values of time series x(t) are defined.

## 2. Information to entropy ratio for signals of chaotic dynamical systems

Let us consider different types of dynamical systems described by

- one-dimensional logistic map [39]

$$x_{i+1} = rx_i(1-x_i), \quad (8)$$

where r is a control parameter;

- two-dimensional Henon map [40]

$$x_{i+1} = 1 - ax_i^2 - by_i, \quad y_{i+1} = x_i. \quad (9)$$

where a and b are control parameters;

- "bursting" map [41]

$$x_{i+1} = (\frac{1}{C} + \mu_i)|x_i|^{-\frac{1}{\gamma}}, \qquad \mu_{i+1} = -\frac{1}{\gamma}(\frac{1}{C} + \mu_i)|x_i|^{-\frac{1}{\gamma}-1}, \quad (10)$$

where $\gamma$ is a fractional part of fractal dimension of a considered physical quantity, C is a parameter similar to compression gain of a fractal signal, $\mu_i$ is a multiplier [41].

Dependence of IER on control parameters calculated by use of Eq. (7) and bifurcation diagrams built with taking into account maximum ($X_{max}$) and minimum ($X_{min}$) local values of the signal described by Eqs. (9), (10) and (11) are shown in Figure 1.

Value of IER decreases in case of appearance of complex cycles corresponding to doubling S2 and quadruplicating S4 of the period, as can be seen in Figure 1(a, b). IER is minimal in case of transition to chaos (Figure 1(c)). Chaotization of a process (decreasing of ordering of a signal) can be also seen from the bifurcation diagram shown in Figure 1(b).

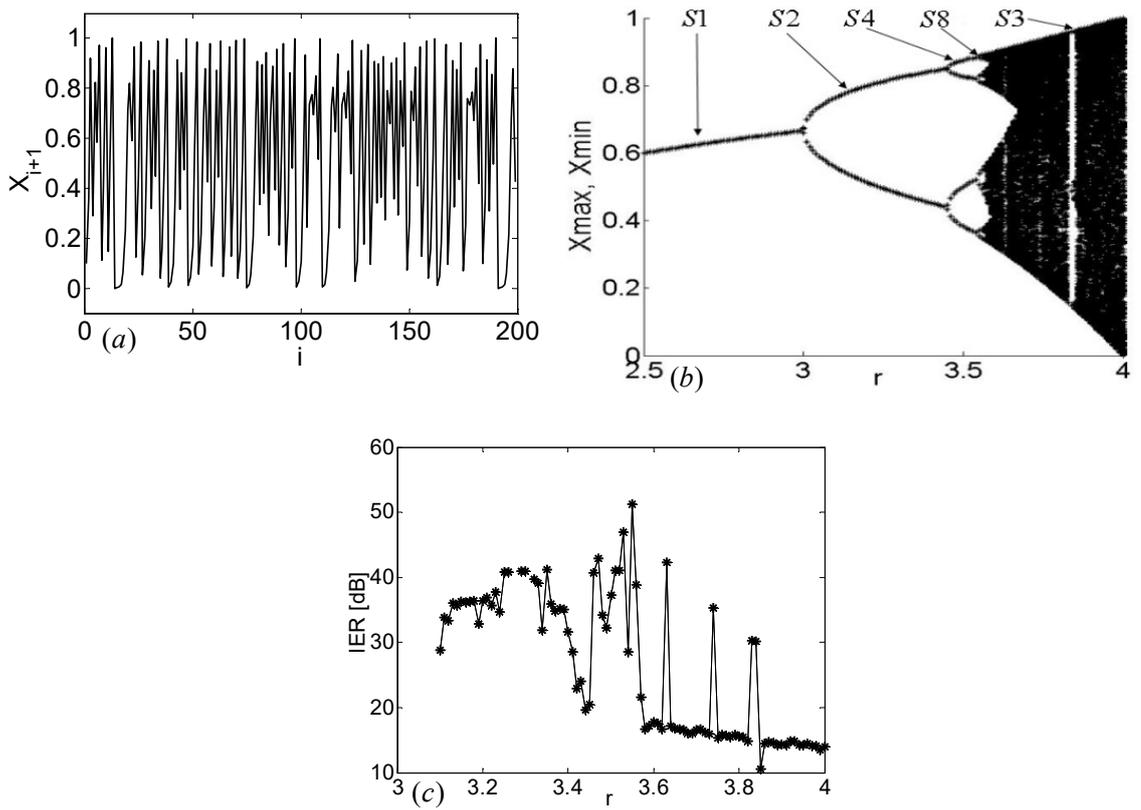

Figure 1 - Time series at r = 4 (a), bifurcation diagram (b) of the logistic map (9) at iteration step δ = 0.01 and dependence of IER on control parameter r (c)

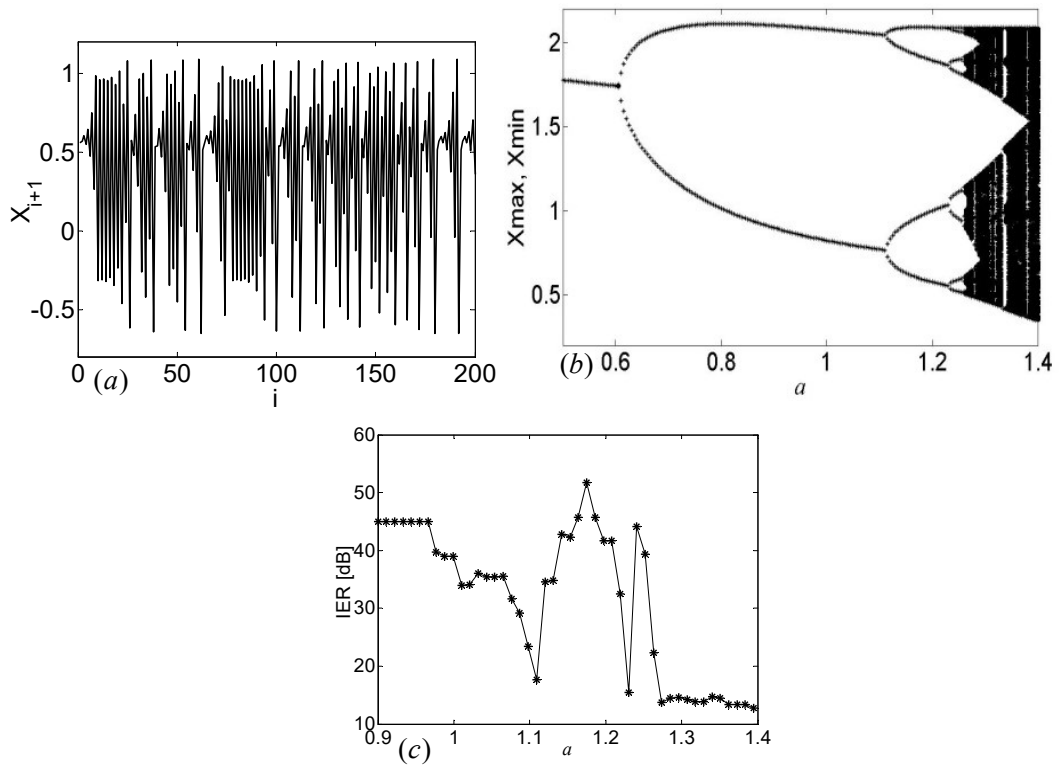

Figure 2 - Time series at a = 1.4, b = 0.1 (a), bifurcation diagram of Henon map at b = 0.1 (b);

δ = 0.01 and the dependence of IER on control parameter a (c)

The same regularities are also observed for two-dimensional Henon map (Figure 2) and "bursting" map (Figure 3). For periodic and quasi-periodic regimes value of IER is greater than for chaotic regimes. Figure 3 shows that peaks of IER can be observed in "windows" of intermittency.

Let us compare values of IER with values of SNR defined by the following relation [4]

$$SNR = 10*\log10(P_S/P_N) = 10*\log10(\sum_{i=1}^{n} S^2[i]/\sum_{i=1}^{n} N^2[i]), \qquad (11)$$

where S and N, $P_S$ and $P_N$ are amplitudes and powers of signal and noise, respectively.

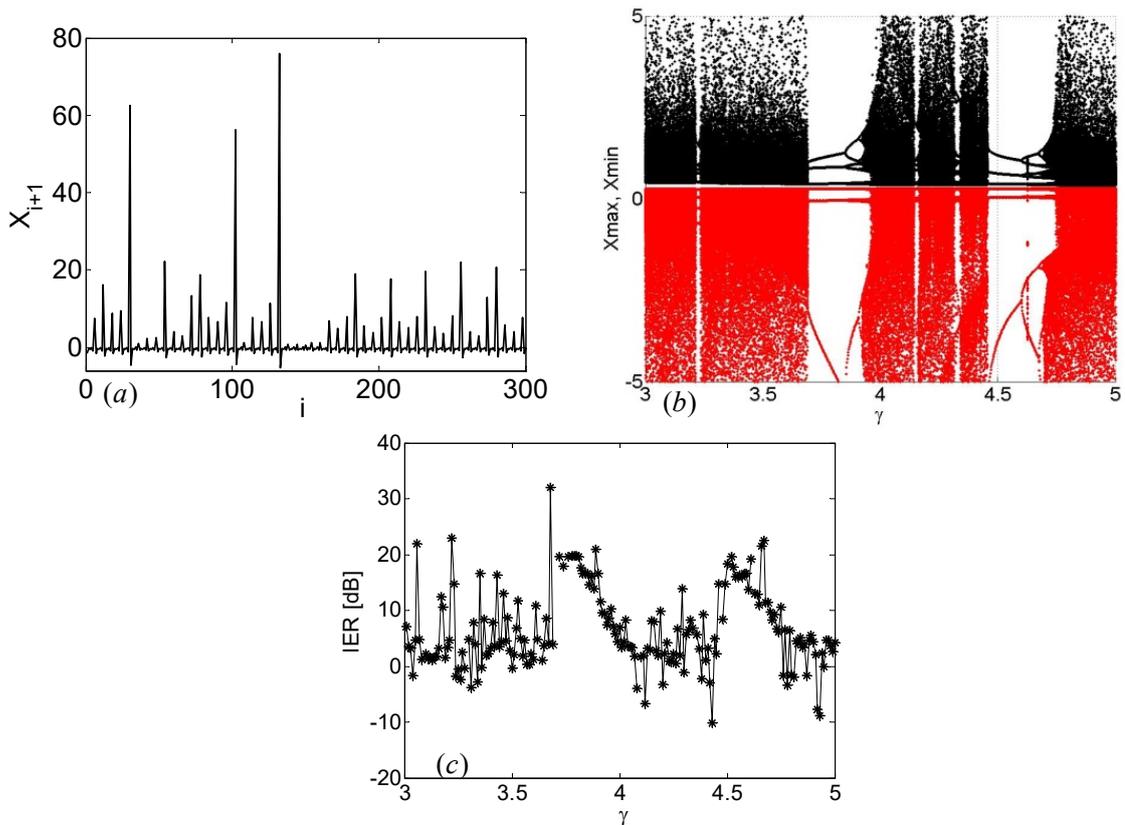

Figure 3 - Time series at γ = 4.4 (a), bifurcation diagram of "bursting" map at C = 2,806 (b); δ = 0.01 and dependence of IER on control parameter γ (c).

For this aim it is necessary to describe a signal and noise separately. We use a sine signal as an informational signal, and signals described by mentioned above maps and Gaussian noise we consider as noise ξ (t). So, we have

$$s(t) = sin(t) + A*\xi(t) \qquad (12)$$

We gradually increase amplitude of noise A and define IER and SNR for different signals (Figure 4). Increasing of noise amplitude A leads to decreasing of SNR for all described above dynamical systems (uncolored marks on the graph). Values of IER defined via Eq. (7) (colored marks) decrease also, but by use of these data we can specify a signal waveshape. Therefore, the new characteristic IER can be used can be used for the description of topological singularities of signals.

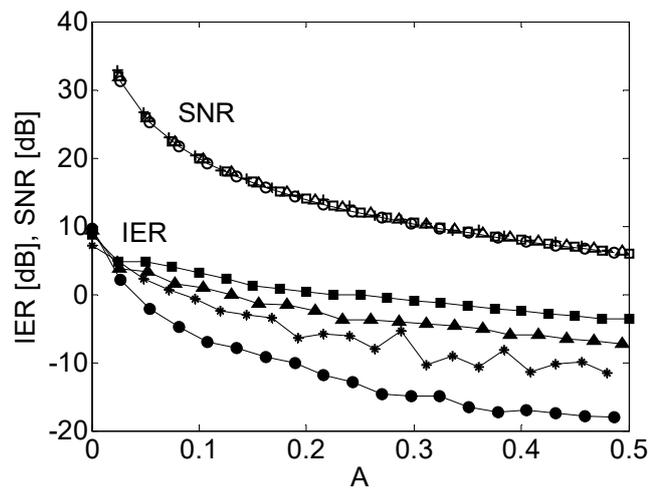

Figure 4 - Dependence of IER (colored marks) and SNR (uncolored marks) at masking of sine signal with noise. □, ■ - logistic map, r = 4; ∆, ▲ - Henon map, a = 1.38, b = 0.1; o, ● - "bursting» map, γ = 4.43, C = 2,806 and + * -White noise ξ (t) with power $P_N$ = 1 W.

## Conclusions

We developed a new universal method for SNR definition on the base of informational-entropic analysis. We suggested the new characteristic (IER) which is the information to entropy ratio. Our approach has the following advantages. As opposed to the well-known methods for SNR definition, IER can be defined even if noise level is unknown. Estimation of IER is possible for short time ranges less than 10 sec in real time.

The new method for SNR estimation via IER based on the theoretical approach which can be used for different applications. We developed algorithms for determination of normalized values for information to entropy ratio.

Results of numeric analysis describing mixtures of regular, chaotic, and stochastic signals demonstrate that using of our method let us to define SNR correctly, and our approach can be used for specification of a signal waveshape.

As known, entropy depends on minimal scale of measurement, and for continuous processes entropy is unlimited. We solved the problem of normalization of entropy by using of information defined via difference between conditional and unconditional entropy normalized on unconditional entropy.